# Exploring Sound *vs* Vibration for Robust Fault Detection on Rotating Machinery

Serkan Kiranyaz, Ozer Can Devecioglu, Amir Alhams, Sadok Sassi, Turker Ince, Onur Avci, and Moncef Gabbouj

*Abstract*— **Robust and real-time detection of faults on rotating machinery has become an ultimate objective for predictive maintenance in various industries. Studies have indicated that around half of all motor failures are attributed to bearing faults alone. Vibration-based Deep Learning (DL) methodologies have become the *de facto* standard for bearing fault detection as they can produce state-of-the-art detection performances under certain conditions. Despite such particular focus on the vibration signal, the utilization of sound, on the other hand, has been neglected whilst only a few studies have been proposed during the last two decades, all of which were based on a conventional ML approach. One major reason is the lack of a benchmark dataset providing a large volume of both vibration and sound data over several working conditions for different machines and sensor locations. In this study, we address this need by presenting the new benchmark *Qatar University Dual-Machine Bearing Fault Benchmark* dataset (QU-DMBF), which encapsulates sound and vibration data from two different motors operating under 1080 working conditions overall. Then we draw the focus on the major limitations and drawbacks of vibration-based fault detection due to numerous installation and operational conditions. Finally, we propose the first DL approach for sound-based fault detection and perform comparative evaluations between the sound and vibration over the QU-DMBF dataset. A wide range of experimental results shows that the sound-based fault detection method is significantly more robust than its vibration-based counterpart, as it is entirely independent of the sensor location, cost-effective (requiring no sensor and sensor maintenance), and can achieve the same level of the best detection performance by its vibration-based counterpart. With this study, the QU-DMBF dataset, the optimized source codes in PyTorch, and comparative evaluations are now publicly shared.**

*Index Terms*—**Operational Neural Networks; Bearing Fault Detection; Machine Health Monitoring.**

## I. INTRODUCTION

Accurate and instantaneous fault detection, especially for rotating machinery, is crucial for motor health monitoring and is required for several industries, such as mass production lines, manufacturing, aerospace, and energy. The most important components of rotating machinery are its bearings as they tend to fail in time which will eventually cause unexpected downtime, high maintenance costs, and even catastrophic accidents if not detected and prevented in advance.

Numerous methodologies have focused on the vibration signal to detect and identify the bearing faults. They can be classified as: model-based methods [1]-[4], traditional signal-processing approaches, [5]-[11], conventional machine learning (ML) and recent deep learning (DL) methods, [12]-[27]. Especially during the last decade, DL-based methods based on the vibration signal have increased tremendously. This is an expected outcome since the vibration signal can easily reveal the track changes over the mechanical behavior of a bearing, and thus, it has become a *de facto* standard in this domain. However, acquiring a reliable and high-quality vibration signal, first of all, requires good-quality sensors which further demands periodic maintenance. This may not only induce significant costs but also pose certain risks of malfunctioning over time. A fundamental problem is that the vibration signal is highly sensitive to the sensor location on the machinery. Take, for example, sample vibration signals acquired from the two machines of the QU-DMBF dataset [27] as shown in Figure 1. Despite some of the sensors being in a close vicinity, the signals from them are entirely different from one another, e.g., see the signals of sensors #2 and #5 from Machine A and sensors #6 and #3 as well as #2, #5, and #4 from Machine B. Therefore, any slight mispositioning of the sensor may result in poor detection performance by the DL method. Moreover, installing such wired sensorial equipment to nearby locations of rotating bearings will lead to certain challenges and operational drawbacks. Finally, vibration signals can easily get corrupted by several background noises such as ambient vibrations, electrical interference of the motor, and sensor variations. Most aforementioned studies ignored such variations and evaluated their proposed method assuming only one or few working condition(s) over the early benchmark datasets with limited vibration data, and a fixed sensor location. In particular, assuming the existence of sufficient fault data for

S. Kiranyaz, is with the Electrical Engineering Department, Qatar University, Doha, Qatar (e-mail: mkiranyaz@qu.edu.qa).
A. Alhams, and S. Sassi are with the Mechanical Engineering Department, Qatar University, Doha, Qatar (e-mail: aa1702913@qu.edu.qa; sadok.sassi@qu.edu.qa)
O. Devecioglu and M. Gabbouj are with the Department of Computing Sciences, Tampere University, Tampere, Finland (e-mail: ozer.devecioglu@tuni.fi, moncef.gabbouj@tuni.fi,).

T. Ince, is with the Electrical and Electronics Engineering Department, Izmir University of Economics, Izmir, Turkey (email:turker.ince@ieu.edu.tr).
O. Avci is with the Department of Civil, Construction and Environmental Engineering, Iowa State University, Ames, IA, USA (email: onur.avci@mail.wvu.edu ) all working conditions to train the fault detector may not be feasible in practice.

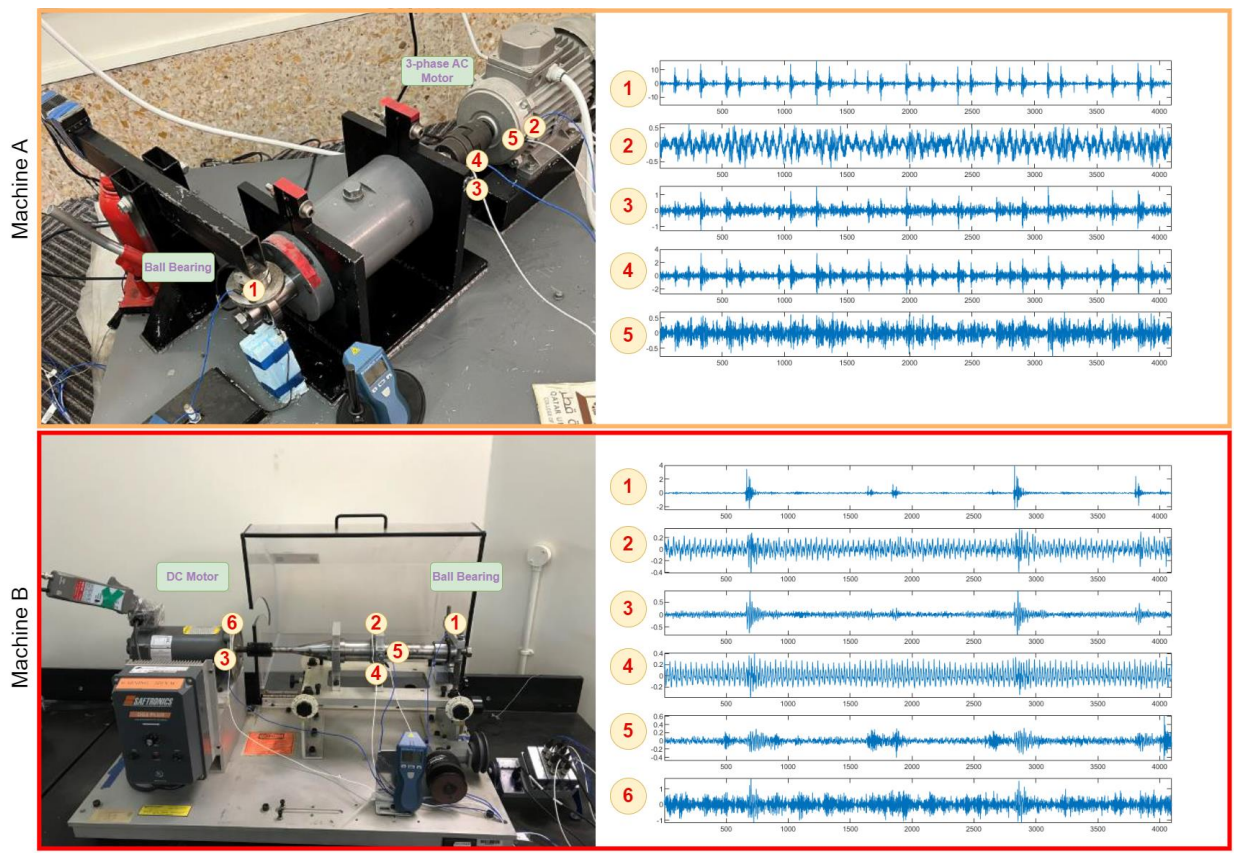

**Figure 1** Samples of *simultaneously* recorded vibration signals over different sensor locations on the QU-DMBF [58].

Conversely, the sound signal has none of the aforementioned issues. For instance, it does not require an additional sensor, and thus, there is no need for any maintenance and installation since a mobile phone can easily acquire sound data from any location, which obviously does not affect the sound signal pattern. Moreover, it is immune from any electrical or sensorial noise. Despite such crucial advantages, only a few sound-based fault detection methods have been proposed [28]-[38], which utilize conventional ML methods such as KNN and SVM classifiers over the manually extracted/selected features. Two possible reasons can be listed: 1) the lack of a benchmark dataset providing a large volume of both vibration and sound data over several working conditions for different motors and sensor locations, and 2) an early study [35] has reported certain limitations of sound signals for identifying defects in gearboxes. It was concluded that the application of sound to gear defect detection is fraught with difficulties, particularly for fault identification. This may have discouraged the researchers from pursuing this direction and drawn the focus rather on the vibration-based approaches. As a result, no DL-based method over the sound signal has ever been proposed to date and comparative evaluations against the vibration counterpart have obviously not yet been reported.

To address all the aforementioned drawbacks, this study proposes the first DL-based fault detection method over the raw sound signal and performs a wide range of experiments for comparative evaluations against its vibration-based counterpart. The QU-DMBF dataset compiled for this purpose is the largest sound/vibration dataset with 13.5 hours of data acquired from two motors (DC and 3-phase AC) with 1080 working conditions. For simplicity and unbiased comparisons, we do not use any Domain Adaptation (DA) methodology [26], [27], [31], and hence, for both signal types, we evaluate their performance over the "unseen" working conditions using a single network model trained over the same data and with an identical experimental setup. In this way, we can test and compare their *robustness* for detecting the "unseen" fault cases, especially when the sensor locations and fault severities for testing differ from the ones used for training.

Self-Organized Operational Neural Networks (Self-ONNs) [25], [27], [47]-[57] are heterogeneous network models with generative neurons that can perform optimal nonlinear operations for each kernel element. Therefore, they can outperform their linear counterparts, the CNNs, in many tasks significantly even with reduced network complexity and depth. In this study, we aim to leverage this superiority to further achieve an elegant computational efficiency for sound-based fault detection. So, a 7-layer 1D Self-ONN with only 80 generative neurons is used for both sound and vibration-based fault detection, and this demonstrates that even with such a compact and lightweight network model, state-of-the-art sound-based fault detection performance can be achieved.

The rest of the article is organized as follows: the exploration methodology with the proposed network model and the QU-

DMBF benchmark dataset are presented in Section II. The fault detection results with detailed comparative evaluations over the QU-DMBF dataset are presented in Section III. Finally, Section IV concludes the paper and suggests topics for future research.

## II. EXPLORATION METHODOLOGY AND QU-DMBF DATASET

### A. 1D Self-Organized Operational Neural Networks

Self-ONNs [25], [27], [47]-[57] were proposed as the superset of the CNNs with a controllable parameter, *Q*, which determines the level of the nonlinearity (degree of the polynomials) of each kernel transformation. When *Q=1* for all neurons in the network, a Self-ONN will reduce to a CNN. Due to the "on-the-fly" generation of the nonlinear nodal operator, the network can create the best possible basis functions so as to achieve the highest learning performance. So, with such optimized nonlinearity and heterogeneity, a Self-ONN can easily surpass an equivalent or even a significantly deeper and more complex CNN.

Each generative neuron can have an arbitrary nodal function, $\Psi$, for each kernel element of each connection. This great flexibility permits the formation of any nodal operator function. For the formation of "any" nodal function, we use the Taylor series approximation as stated in:

$$\Psi(\mathbf{w}, y) = w_0 + w_1 y + w_2 y^2 + \cdots + w_Q y^Q \quad (1)$$

Any other function approximation technique (e.g., Fourier series or DWT) can also be used, but such techniques will also require the computation of certain basis (nonlinear) computationally demanding functions. However, as formulated in [40] and [41], the generative neurons can be turned into purely a set of convolutions, yielding a great computational advantage.

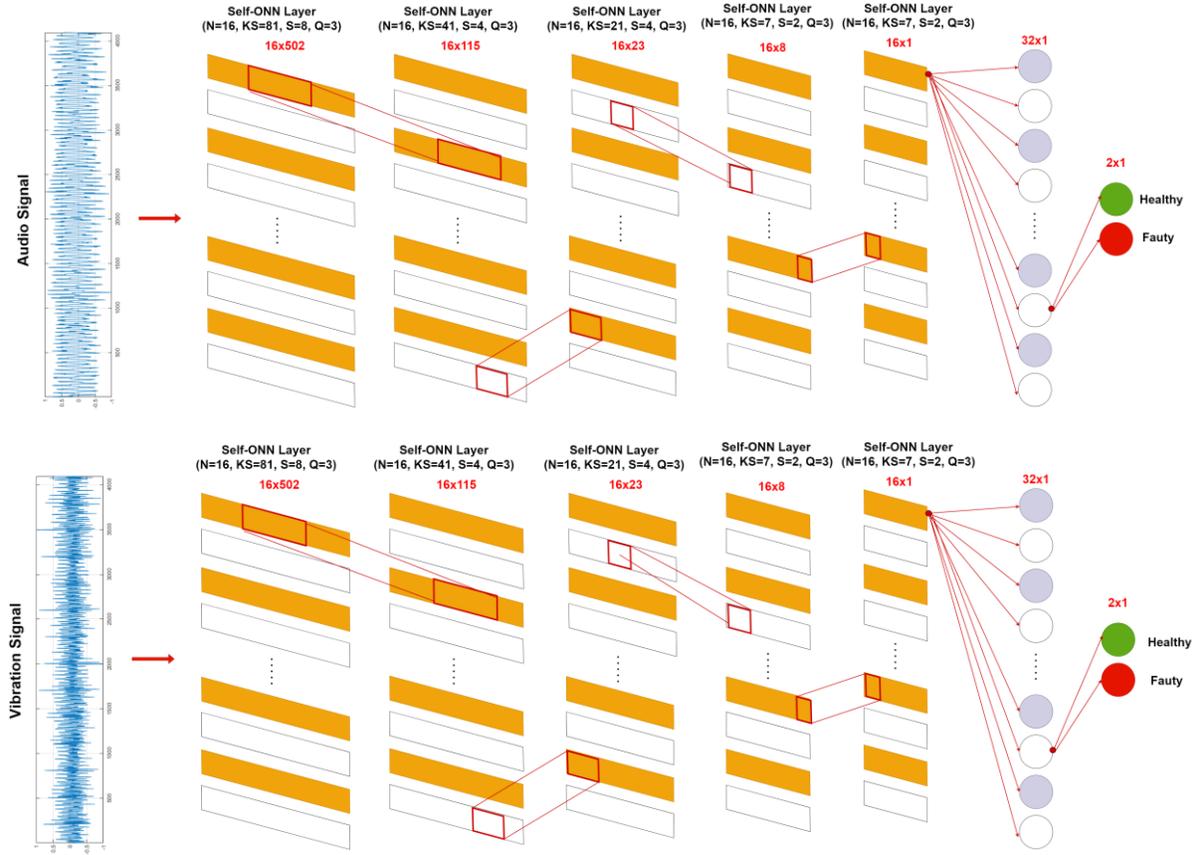

**Figure 2:** Identical 7-layer Self-ONN models used for sound (top) and vibration-based (bottom) fault detection. The number of neurons (N), 1D kernel size (KS), stride (S), and Q values are also presented for each layer.

Let $x_{ik}^l \in \mathbb{R}^M$ be the contribution of the $i^{th}$ neuron at the $(l-1)^{th}$ layer to the input map of the $l^{th}$ layer. Therefore, it can be expressed as,

$$\widetilde{x_{ik}^l}(m) = \sum_{r=0}^{K-1}\sum_{q=1}^{Q} w_{ik}^{l(Q)}(r,q)\left(y_i^{l-1}(m+r)\right)^q \quad (2)$$

where $y_i^{l-1} \in \mathbb{R}^M$ is the output map of the $i^{th}$ neuron at the $(l-1)^{th}$ layer, $w_{ik}^{l(Q)}$ is a learnable kernel of the network, which is a $K \times Q$ matrix, i.e., $w_{ik}^{l(Q)} \in \mathbb{R}^{K \times Q}$, formed as, $w_{ik}^{l(Q)}(r) = [w_{ik}^{l(Q)}(r,1), w_{ik}^{l(Q)}(r,2), \ldots, w_{ik}^{l(Q)}(Q)]$. By the commutativity of the summation operations in (2), one can alternatively express:

$$\widetilde{x_{ik}^l}(m) = \sum_{q=1}^{Q} \sum_{r=0}^{K-1} w_{ik}^{l(Q)}(r, q-1) y_i^{l-1}(m+r)^q \quad (3)$$

One can simplify this expression as follows:

$$\widetilde{x_{ik}^l} = \sum_{q=1}^{Q} Conv1D\left(w_{ik}^{l(Q)}, \left(y_i^{l-1}\right)^q\right) \quad (4)$$

Eq. (4) simply shows that the kernel operations can indeed be executed by performing $Q$ number of 1D convolutions. Finally, the input map of this neuron can be composed as,

$$x_k^l = b_k^l + \sum_{i=0}^{N_{l-1}} x_{ik}^l \quad (5)$$

where $b_k^l$ is the bias associated with this neuron. Passing the input map through the activation function will generate the output map, $y_k^l = f(x_k^l)$, which will then contribute to the input maps of the neurons on the next layer and so on. For a parallel processing implementation, in [40], [42], and [48], the raw-vectorized formulations of both FP and BP are presented.

Figure 2 shows the identical 1D Self-ONN models for both sound and vibration-based fault detection. Each model contains five operational layers with a total of 80 generative neurons, one dense layer with 32 generative perceptrons, and an output layer with two generative perceptrons. For each layer, $Q=3$ and *tanh* activation functions are used. The kernel sizes at the consecutive operational layers are set as: 81, 41, 21, 7, and 7, respectively.

Both vibration and sound data are framed into one-second segments without any overlapping. The sampling frequency of both signals is 4096 Hz, and thus, each segment has $m = 4096$ samples. For a strict magnitude invariance during the training (BP) and forward propagation (FP) on the 1D Self-ONN, each segment is further normalized as follows:

$$X_N(i) = \frac{2(X(i) - X_{min})}{X_{max} - X_{min}} - 1 \quad (6)$$

where $X(i)$ is the $i^{th}$ original sample amplitude in the segment, $X_N(i)$ is the $i^{th}$ sample amplitude of the normalized segment, $X_{min}$ and $X_{max}$ are the minimum and maximum amplitudes within the segment, respectively. This will scale the segment linearly in the range of [-1 1], where $X_{min} \rightarrow -1$ and $X_{max} \rightarrow 1$. The proposed 1D Self-ONN model receives the normalized segment samples as input channel and produces the binary output (class) vector.

*B. Qatar University Dual-Machine Bearing Fault Benchmark Dataset: QU-DMBF*

Researchers from Qatar University created the benchmark dataset for this study using two distinct electric machines (Machine A and Machine B). Figure 1 depicts the experimental setup, including the positioning of the sensors and the installation of two machines. A three-phase AC motor, two double-row bearings, and a shaft with a maximum speed of 2840 RPM make up Machine-A's configuration. Two different loads (0.18 kN and 0.23 kN ) were applied using an arm that is attached to a hydraulic jack from one end and to a plate from the other end. This plate is used to apply the load on the outer race of the bearing. Mounted on the bearing housing are PCB accelerometers (352C33 high-sensitivity Quartz ICP). The machine measures 100x100x40 cm and weighs 180 kg. The following is the outline of Machine-A's working conditions:
- There are 19 distinct bearing configurations: 1 healthy case, and 18 faulty cases that consist of 9 outer ring defects and 9 inner ring defects. The range of defect sizes varies from 0.35 mm to 2.35 mm.
- There are 5 distinct locations for accelerometers: two radial directions and three horizontal locations.
- There are two distinct force levels: 0.18 kN and 0.23 kN.
- Three distinct RPM ranges: 480, 680, and 1010 RPM.

For a healthy bearing, we recorded data for 270 seconds of operation, and for a damaged bearing case, we recorded 30 seconds. 30 x 18 x 5 x 2 x 3 = 16,200 seconds (4.5 hours) of defect data, and 270 x 5 x 2 x 3 = 8,100 seconds (2.25 hours) of healthy data are the totals that come from this.

Machine B, on the other hand, has a DC motor, two single-row bearings, and a shaft that rotates at a varying speed of around 2000 RPM (max. 2500rpm). A constant load of 0.18kN was applied to the bearing by tightening 2 bolts at the two ends of a rectangular thick plate to let it press on the outer race of the bearing. Mounted on the bearing housing are PCB accelerometers (353B33 high-sensitivity Quartz ICP). The entire machine is 59 kg in weight and has dimensions of 100x63x53 cm. The following are the different working conditions for Machine B:
- 9 bearing configurations with an outer ring defect, nine with an inner ring defect, and one healthy bearing configuration. The range of fault sizes varies from 0.35 mm to 2.35 mm.
- There are 6 distinct locations for accelerometers.
- A constant force (load) of 0.18 kN.
- Five distinct RPM ranges: 240, 360, 480, 700, and 1020.

For each working condition of a healthy bearing, 270 seconds of vibration/sound data are recorded. Thus, the total duration of the vibration data for a healthy bearing is 270 x 6 x 1 x 5 = 8,100 seconds (2.25 hours). In a similar manner, 30 seconds of sound and vibration data are recorded for every working condition of a faulty bearing. The faulty to healthy data ratio is 2:1, and the total time is 30 x 18 x 6 x 1 x 5 = 16,200 seconds (4.5 hours). Consequently, the entire duration of the dataset on machine B is 24,300 seconds (6.75 hours). For all working conditions in both machines, the sound was also simultaneously recorded with the same sampling frequency as the vibration data.

*C. Experimental Setup*

A recent study [27] has shown that the most reliable vibration data for fault detection is acquired from the closest accelerometer to the bearings, i.e., sensor #1 for both machines, as shown in Figure 1. So, we have selected a part of the data of this accelerometer for training both fault detectors and use the rest of the data and the data of other sensors for testing. In particular, the sound and vibration data of sensor #1 acquired from the two smallest defects (0.35mm and 0.5mm) are used for training, which corresponds to 124 fault segments. This corresponds to only 4.44% and 3.37% of the machine-A and machine-B fault data used for training, respectively. The same number of healthy segments is also used to compose an isolated training data partition to evaluate the robustness of both fault detectors against the variations in sensor locations, fault severities, speed, and load over both machines.

For the Back-Propagation (BP) training, the Adam optimizer is used with the initial learning factor, $\varepsilon=10^{-4}$, and the Mean-Squared-Error (MSE) as the loss function. %20 of the training data is spared as validation to select the best Self-ONN model for testing. We implemented both fault detector networks using the FastONN library [42] based on PyTorch.

Commonly used performance metrics, Precision ($P$), Recall ($R$), F1-Score ($F1$), and Accuracy ($Acc$) are computed to evaluate the fault detection performances. The calculation of True Positives (TP), False Negatives (FN), and False Positives (FP) are obtained per vibration/sound segment classification in the test set. Accordingly, these performance metrics can be expressed as follows:

$$P = \frac{TP}{TP+FP}, \quad R = \frac{TP}{TP+FN}$$
$$F1 = \frac{2PR}{P+R}, \quad Acc = \frac{TP+TN}{TP+TN+FP+FN} \quad (7)$$

## III. EXPERIMENTAL RESULTS

In the next subsection, we shall perform an extended set of comparative evaluations between sound and vibration-based fault detection. In Section III.B, the computational complexity analysis of the proposed network model will be examined in-depth.

### A. Results

Once the fault detectors of both sound and vibration data are trained over the training data (a fraction of healthy and faulty data acquired by sensor #1), they are tested over the test data of each machine in the QU-DMBF dataset. For each machine, average fault detection performances over each sensor are individually computed using the aforementioned standard metrics and presented in Table 1 and Table 2. As the sensor location does not matter for sound, its fault detection results are presented in the first row of the tables.

Table 1: The fault detection performances per sensor over sound and vibration data for machine A.

| Test Sensor # | Accuracy | Precision | Recall | F1-Score |
|---|---|---|---|---|
| Sound @ Sensors 1-5 | 99.38 | 99.77 | 99.38 | 99.58 |
| Vibration @ Sensor #1 | 99.75 | 99.77 | 99.88 | 99.83 |
| Vibration @ Sensor #2 | 62.15 | 85.93 | 57.81 | 69.12 |
| Vibration @ Sensor #3 | 68.53 | 99.40 | 57.39 | 72.77 |
| Vibration @ Sensor #4 | 89.34 | 93.74 | 91.57 | 92.64 |
| Vibration @ Sensor #5 | 48.36 | 99.61 | 29.64 | 45.68 |

Table 2: The fault detection performances per sensor over sound and vibration data for machine B.

| Test Sensor # | Accuracy | Precision | Recall | F1-Score |
|---|---|---|---|---|
| Sound @ Sensors 1-6 | 97.07 | 97.95 | 98.04 | 98.00 |
| Vibration @ Sensor #1 | 96.80 | 99.81 | 95.80 | 97.76 |
| Vibration @ Sensor #2 | 58.73 | 99.89 | 43.56 | 60.67 |
| Vibration @ Sensor #3 | 74.16 | 99.86 | 64.74 | 78.55 |
| Vibration @ Sensor #4 | 53.31 | 99.49 | 36.29 | 53.18 |
| Vibration @ Sensor #5 | 56.85 | 99.66 | 41.09 | 58.19 |
| Vibration @ Sensor #6 | 44.96 | 72.63 | 39.60 | 51.25 |

Several important observations can be made regarding the fault detection results presented in Table 1 and Table 2. First of all, almost identical fault detection performances are obtained by sound and vibration-based fault detectors, but only when the vibration detector is tested on the same sensor data used for its training (sensor #1). In this case, the proposed 1D Self-ONN model has achieved remarkable performance levels between 97.6% to 99.83% F1 scores, on both machines A and B, respectively. However, the performance of vibration-based fault detection significantly deteriorates when the sensor location is altered, even for a slight change. Performance levels around 50% (highlighted in red in both tables) can be observed on both machines, which basically indicates a detection failure since 50% accuracy (or precision, or recall, or F1) is a bottom-line performance level for a binary classification problem. It is worth noting that there is a significant performance gap even over the results obtained from the data of the sensors which are quite close to each other, e.g., see the accuracies obtained for sensor pairs, #2 - #5 and #3 - #4 on Machine A. Such high variations from nearby sensors may also indicate that the mounting of the sensors to the surface of the machines differ from each other and this, in turn, alters the acquired vibration signals significantly as witnessed in Figure 1. Besides the challenges of installing accelerometers over rotating machinery, this further indicates how sensitive fault detection can be with respect to the sensor locations and mounting. On the other hand, sound-based detection has none of such drawbacks, sensitivities, or limitations, as the results on both machines show that it can always achieve a high detection performance with such a lightweight network model trained over a minority of the fault data (<5% of the fault data).

### B. Computational Complexity Analysis

This section presents the computational complexity analysis by measuring the inference time, network size, and total number of parameters (PARs) of the proposed Self-ONN configuration. Comprehensive formulations of the PARs for Self-ONNs can be found in [42]. All trials were conducted using a 2.2 GHz Intel Core i7 PC equipped with an NVIDIA GeForce RTX 3080 graphics card and 16 GB of RAM. For the fault detectors, PyTorch and the FastONN library [42] were utilized. The proposed 7-layer Self-ONN model contains 377K parameters in total. For a single CPU implementation, the forward propagation (FP) time for classifying a one-second segment (sound or vibration) is 4 msec. With a single CPU, this demonstrates that the proposed sound-based fault detector can work 250 times faster than the real-time requirements. This indicates that the proposed approach can be used for a real-time motor health monitoring implementation, even as a typical mobile phone application.

### IV. CONCLUSIONS

With the recent advances in DL models, vibration-based fault detection has become the *de facto* standard for rotating machinery. This study explores the use of sound *vs* vibration in a DL method for fault detection in terms of detection accuracy and robustness over the largest benchmark dataset, QU-DMBF ever composed with 1080 working conditions. The significant and novel contributions of this study can be summarized as follows:

- This is a pioneer study that proposes a DL approach for sound-based bearing fault detection and makes comparative evaluations against its vibration-based counterpart.
- To achieve the objectives of this study, the QU-DMBF benchmark dataset is compiled to obtain both sound and vibration data simultaneously for 1080 working conditions. The QU-DMBF dataset, our results, and the optimized PyTorch implementation of the proposed sound-based detection approach are now publicly shared with the research community [58].
- The proposed sound-based fault detector based on a compact model of 1D Self-ONN achieves state-of-the-art detection performance levels despite only the minority of the sound data being used for training a compact Self-ONN model and tested over a data partition with a large number of "unseen" working conditions.
- An extended set of comparative evaluations has demonstrated that the sound-based detection performance can match the *best* vibration-based counterpart, which is observed only when the detector model is trained and tested over the same sensor data.
- Another important observation of this study is that the vibration-based fault detectors have a high sensitivity to the sensor location, i.e., a slight relocation of the sensor can cause a significant deterioration in the detection performance. This is a crucial advantage of the sound-based detectors that are invariant to such changes.

In brief, the proposed method makes motor health monitoring significantly more robust, practical, cheaper, and accessible. Thus, it has the potential to make a crucial impact on the other related fields of health monitoring and predictive maintenance, e.g., mechanical fault detection on vehicles or other engines used for transportation.

Future research will focus on sound-based fault identification and localization in order to determine the exact nature of each fault, predict its severity, and identify its location for a full-scale motor health monitoring.

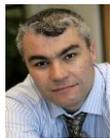 **Serkan Kiranyaz** works as a Professor in Qatar University, Doha, Qatar. He published two books, 7 book chapters, 10 patents/applications, more than 100 journal articles in several IEEE Transactions and other high impact journals, and more than 120 papers in international conferences. His principal research field is machine learning and signal processing. He made significant contributions on bio-signal analysis, classification and segmentation, computer vision with applications to recognition, classification, multimedia retrieval, evolving systems and evolutionary machine learning, swarm intelligence and evolutionary optimization.

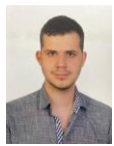 **Ozer Can Devecioglu** was born in Turkey, in 1996. He received the B.S. and M. S. degrees from the Izmir University of Economics, Izmir, Turkey, in 2019 and 2022 respectively. He is currently working as a researcher at Tampere University, Tampere, Finland. His research interests include machine learning, artificial neural networks, image processing, and signal processing.

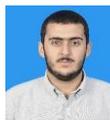 **Amir Alhams** received the bachelor's degree in mechanical engineering from Qatar University in 2021, where he is currently pursuing his master's degree in mechanical engineering. His research interests include Semi-Active Vibration Control of Vehicles, Condition Monitoring, and Diagnosis of Rotating Machine Faults using Artificial Intelligence.

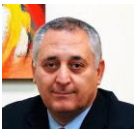 **Dr. Sadok Sassi** is a Professor of Mechanical Engineering at Qatar University. He received his Doctorate in Applied Mechanics from the Polytechnic of Montreal (Montreal, Canada) in 1994. His current research interests include Mechanical Vibration, Stability of Structures, Condition Monitoring of Rotating Machinery, Troubleshooting and Diagnosis of Damaged Machines Components, Numerical Simulation and Experimental Testing of Rotating Mechanical Systems, and Development of Smart-Material-Based Intelligent Dampers.

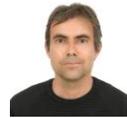 **Turker Ince** received the B.S. degree from the Bilkent University, Ankara, Turkey, in 1994, the M.S. degree from the Middle East Technical University, Ankara, Turkey, in 1996, and the Ph.D. degree from the University of Massachusetts, Amherst (UMass- Amherst), in 2001 all in electrical engineering. From 1996 to 2001, he was a Research Assistant at the Microwave Remote Sensing Laboratory, UMass-Amherst. He worked as a Design Engineer at Aware, Inc., Boston, from 2001 to 2004, and at Texas Instruments, Inc., Dallas, from 2004 to 2006. In 2006, he joined the faculty of Engineering at Izmir University of Economics, Turkey, where he is currently a Professor in Electrical & Electronics Engineering department. His research interests are in the areas of machine learning, signal processing, biomedical engineering, predictive analytics, and remote sensing.

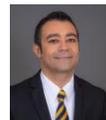 **Dr. Onur Avci** is a faculty member at West Virginia University's Civil and Environmental Engineering department. After receiving M.S. and Ph.D. degrees at Virginia Tech, he worked at major structural engineering firms in the United States as a registered Professional Engineer. During his industry design experience, he was involved in the analysis, design, renovation, and demolition of more than 5 million m2 of structural space. In his research, he focuses on structural dynamics and Machine Learning applications in structural engineering. During his academic experience, he has attracted more than $3.7 million in research grants.

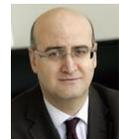 **Moncef Gabbouj** is Professor at the Department of Computing Sciences, Tampere University, Finland. He was an Academy of Finland Professor. His research interests include Big Data analytics, multimedia analysis, artificial intelligence, machine learning, pattern recognition, nonlinear signal processing, video processing, and coding. Dr. Gabbouj is a Fellow of the IEEE and Asia-Pacific Artificial Intelligence Association. He is member of the Academia Europaea, the Finnish Academy of Science and Letters and the Finnish Academy of Engineering Sciences.